\documentclass[twoside,slac_one]{revtex4}
\usepackage{graphicx}
\usepackage{fancyhdr}
\usepackage{amsmath} 
\usepackage{bm}
\usepackage{amsxtra}
\usepackage{amssymb}
\usepackage{amsthm}
\usepackage{latexsym}
\usepackage{lscape}

\begin{document}

\title{KATRIN: an experiment to determine the neutrino mass}

%

\author{F.M. Fr\"ankle}
\affiliation{Department of Physics and Astronomy, University of North Carolina at Chapel Hill, Chapel Hill, NC, USA}

\begin{abstract}
The \textbf{KA}rlsruhe \textbf{TRI}tium \textbf{N}eutrino (KATRIN) experiment is a next generation, model independent, large scale experiment to determine the neutrino mass by investigating the kinematics of tritium beta-decay with a sensitivity of 200~meV/c$^2$. The measurement setup consists of a high luminosity windowless gaseous molecular tritium source (WGTS), a differential and cryogenic pumped electron transport and tritium retention section, a tandem spectrometer section (pre-spectrometer and main spectrometer) for energy analysis, followed by a detector system for counting transmitted beta-decay electrons. To achieve the desired sensitivity, the WGTS, in which tritium decays with an activity of about 10$^{11}$~Bq, needs to be stable on the 0.1~\% level in injection pressure and temperature at an absolute value of about 30~K. With the capability to create an axial magnetic field of 3.6~T the WGTS is going to be one of the world's most complex superconducting magnet and cryostat systems. The main spectrometer (length 24~m, diameter 10~m), which works as a retarding electrostatic spectrometer, will have an energy resolution of 0.93~eV at 18.6~keV. For the precise energy analysis at the tritium endpoint, a retarding potential of -18.6~kV is needed with 1~ppm stability. To reach the background level needed to achieve the sensitivity, it will be operated at a pressure of 10$^{-11}$~mbar. This article will give an overview of the KATRIN experiment and its current status.
\end{abstract}

\maketitle

\thispagestyle{fancy}


\section{Introduction}

\begin{figure*}[hb]
	\centering
	\includegraphics[width=135mm]{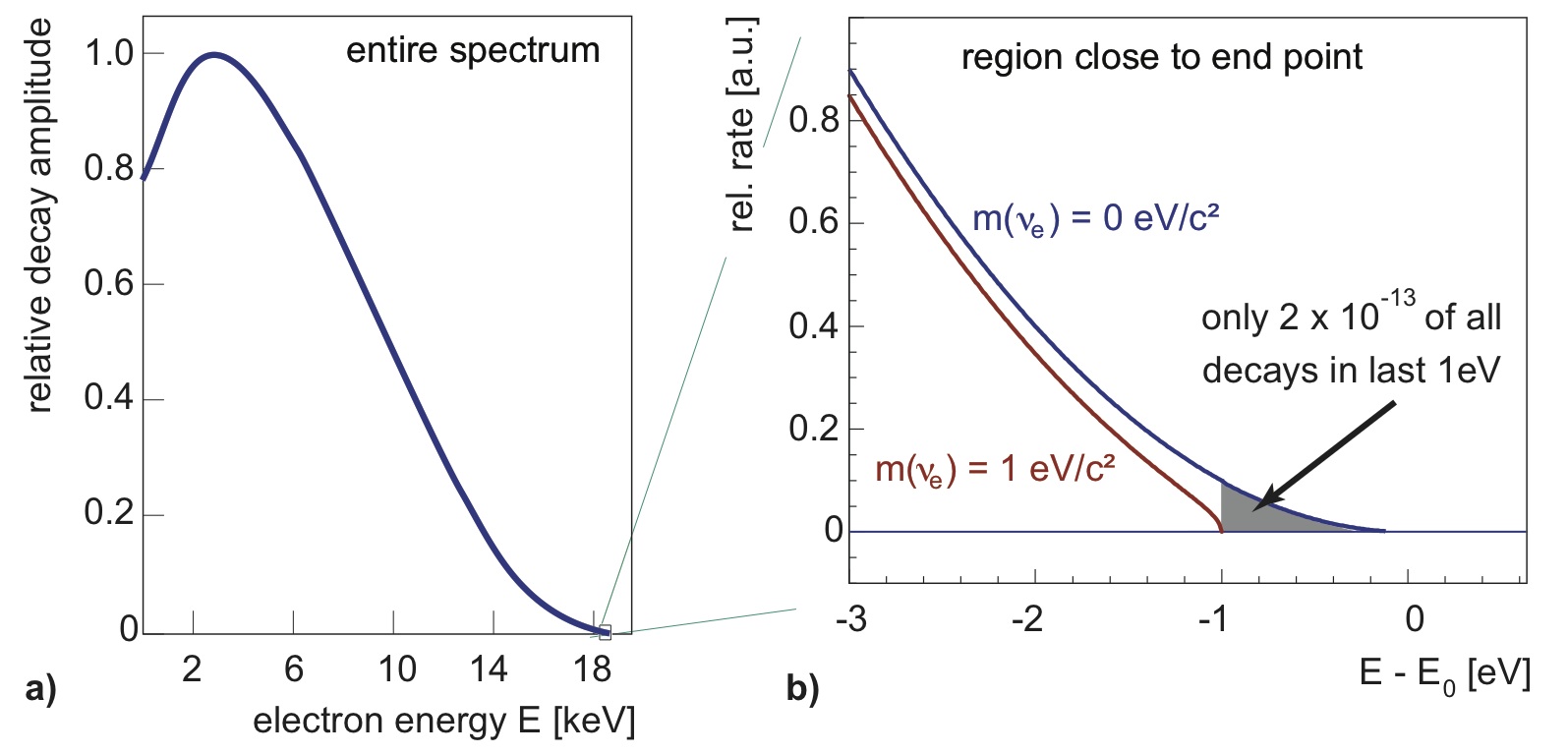}
	\caption{\textbf{a)} Tritium $\beta$-spectrum, \textbf{b)} region close to end point, two scenarios for the neutrino mass are shown ($m^{2 \mathsf{(eff)}}_{\bar{\nu_e}}$~=~0 and 1~eV/c$^2$)} \label{fig:betaspec}
\end{figure*}

Neutrinos have broad implications for particle physics and cosmology in particular with regard to their absolute mass scale. Due to their large abundance (the neutrino density of the universe is 336 $\nu$'s cm$^{-3}$ for all three flavors) they affect the evolution of gravitational clustering depending on their mass. Neutrino oscillation experiments imply that neutrinos are massive and set a lower limit of 40~meV/c$^2$ \cite{nak10} on the neutrino mass. Experiments investigating the kinematics of $\beta$-decay on the other hand set an upper limit on the neutrino mass of 2.3~eV/c$^2$ \cite{kra05, lob99}. The attempts to determine the absolute neutrino mass from cosmological data strongly depend on the chosen data set and model. The KATRIN experiment \cite{kat05} uses a model independent method to determine the neutrino mass ($m^{2 \mathsf{(eff)}}_{\bar{\nu_e}}$) by investigating the kinematics of tritium $\beta$-decay. In nuclear $\beta$-decay a neutron in the atomic nucleus decays into a proton, thereby emitting an electron ($e^-$) and an electron anti-neutrino ($\overline{\nu}_e$). The energy released in the decay is stored between the $e^-$ and $\overline{\nu}_e$ in a statistical way. The energy spectra of the electron is given by the well known Fermi theory of $\beta$-decay \cite{fer34}:

\begin{equation}
	\frac{dN}{dE} \propto p(E+m_ec^2)(E_0-E) \sqrt{(E_0-E)^2 -  m_{\overline{\nu}_e}^{2  \mathsf{(eff)}} c^4}
	\label{equ:fermibeta}
\end{equation}

with the the electron energy $E$, the endpoint energy $E_0$ , the electron mass $m_e$ and the neutrino mass $m^{2 \mathsf{(eff)}}_{\bar{\nu_e}}$. As one can see in equation \ref{equ:fermibeta}, it is the square of the neutrino mass $m_{\overline{\nu}_e}^2$ that enters as a parameter. Its effect on the shape of the spectrum is significant only in a very narrow region close to $E_0$. Figure \ref{fig:betaspec}b) shows the energy spectrum of tritium $\beta$-electrons for the case of $m^{2 \mathsf{(eff)}}_{\bar{\nu_e}}$ = 0~eV/c$^2$ and $m^{2 \mathsf{(eff)}}_{\bar{\nu_e}}$ = 1~eV/c$^2$, underlining the impoertance of the end point regions, where non-relativistic neutrinos are emitted. The KATRIN experiment aims to determine the observable $m_{\overline{\nu}_e}^2$ by measuring the shape of the $\beta$-spectrum close to its endpoint. KATRIN uses molecular tritium as $\beta$-emitter because it has several advantages compared to other isotopes:

\begin{itemize}
	\item \textbf{low endpoint energy}: Tritium (T) has one of the lowest endpoint energies ($E_0$~=~18.59~keV \cite{nag06}) of all $\beta$-active isotopes. The low endpoint energy maximizes the event rate close to the endpoint region.
	\item \textbf{high specific activity}: The specific activity of tritium is 3.6~$\cdot$~10$^{14}$~Bq/g which allows to build a source of high luminosity. This is an important cornerstone for KATRIN in order to measure the energy spectra close to $E_0$ with sufficient statistics (only $<$ 10$^{-11}$ of all tritium $\beta$-decays happen within a few eV below $E_0$, see figure \ref{fig:betaspec}b).
	\item \textbf{super-allowed transition}: The nuclear transition of tritium $\beta$-decay is super-allowed, hence there are no energy dependent corrections of the nuclear matrix elements.
	
\end{itemize}

With a sensitivity of 200~meV/c$^2$ KATRIN has the potential to probe the entire quasi-degenerated $\nu$-mass scale and the complete mass range of  $m^{2 \mathsf{(eff)}}_{\bar{\nu_e}}$ where neutrinos have a significant influence on structure formation in the universe.

\section{Experimental setup and status}

\begin{figure*}[hb]
	\centering
	\includegraphics[width=135mm]{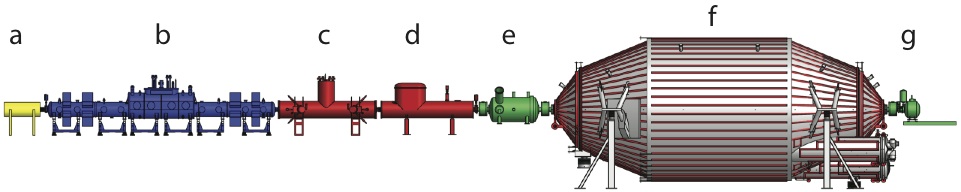}
	\caption{Schematic overview of the KATRIN experimental setup, \textbf{a} rear section, \textbf{b} windowless gaseous molecular tritium source (WGTS), \textbf{c} differential pumping section (DPS), \textbf{d} cryogenic pumping section (CPS), \textbf{e} pre-spectrometer, \textbf{f} main spectrometer, \textbf{g} detector system. The overall setup has a length of about 70 m} \label{fig:katrin}
\end{figure*}

The KATRIN experiment is currently under construction at the Karlsruhe Institute of Technology (KIT) in Germany. The measurement setup (see figure \ref{fig:katrin}) has an overall dimension of $\approx$~70~m and consists of a high luminosity windowless gaseous molecular tritium source WGTS (\textbf{b}) whose activity is monitored by the rear section (\textbf{a}), a differential (\textbf{c}) and cryogenic (\textbf{d}) tritium pumping and retention section, a tandem spectrometer section with a pre-spectrometer (\textbf{e}) and a main spectrometer (\textbf{f}) for energy analysis, which is finally followed by a detector system (\textbf{g}) for counting the transmitted $\beta$-decay electrons. In the following each component is described in more detail. 

\subsection{Windowless gaseous tritium source (WGTS)}

The tritium source consists of a 10~m long stainless steel beam tube with a diameter of 90~mm which is operated at a base temperature of 30~K, maintained by a dedicated two phase Neon cooling system. Tritium molecules (T$_2$) are injected into the centre of the tube (injection pressure 3.35~$\cdot$~10$^{-3}$~mbar). While diffusing to the ends of the beam tube tritium decays with a rate of 10$^{11}$~Bq thus providing a sufficient number of $\beta$-electrons close to $E_0$ (see figure \ref{fig:betaspec}). The activity of the source will be continuously monitored by the rear section. The tritium at the end of the WGTS beam tube is collected via turbomolecular pumps (TMPs) and recirculated via an "inner loop" \cite{stu10}, which removes contaminants (particularly $^3$He) and is capable to process a throughput of 40~g T$_2$ per day. Accordingly, a T$_2$ pressure gradient with a column density of 5~$\cdot$~10$^{17}$~T$_2$/cm$^2$ is created within the source. In order to keep the systematic errors of the source on a level comparable to the statistical errors, the column density needs to be stable on a 0.1~\% level. This implies stringent requirements for the temperature stability and homogeneity of the source of $\Delta T\leq \pm$30~mK as well as of 0.1~\% for the injection pressure.

The $\beta$-electrons created inside the WGTS are guided via an axial magnetic field of $B_S$~=~3.6~T towards the spectrometers. The start angle $\Theta$\footnote{$\Theta$ is defined as the angle between electron momentum $\vec{p}$ and magnetic field $\vec{B}$.} defines the path length of electrons in the WGTS, meaning that electrons have different probabilities for inelastic scattering on T$_2$. In order to minimize the number of electrons undergoing inelastic collisions for the neutrino mass measurements, the maximum accepted start angle is limited to $\Theta_{max}$~=~51$^{\circ}$ via a maximum magnetic field of $B_{max}$~=~6~T (pinch magnet) which reflects electrons with $\Theta > \Theta_{max}$ due to magnetic mirror effect. The diameter of the WGTS defines the magnetic flux $\Phi$ that needs to be transported from the WGTS to the detector system. With an effective source area of $A_S$~=~53~cm$^2$ and a magnetic field of $B_S$~=~3.6~T, $\Phi$ can be calculated to be $\Phi$~=~191~Tcm$^2$ with

\begin{equation}
	\Phi = \int{B(r) rdr \approx A_S B_S}
	\label{equ:fluxtube}
\end{equation}

The integral is given in cylindrical coordinates, referring to the axial symmetric setup of the KATRIN experiment. For a homogeneous magnetic field the integral simplifies to a product of magnetic field and area.

Measurements with the WGTS demonstrator - a prototype to test the cooling concept of the WGTS - showed that the requirements for temperature stability of the source tube can be met. The construction of the WGTS will start in 2012.

\subsection{Transport and tritium retention section}

The task of the beam element DPS2-F is to reduce the T$_2$ partial pressure by a factor of $>$~10$^5$ and to guide the $\beta$-electrons via a strong magnetic field of 5.6~T. The beam tube has four bends of 20$^{\circ}$ to avoid beaming of T$_2$ molecules towards the spectrometers and to increase the effective pumping speed of the turbomolecular pumps attached there. The DPS2-F arrived on site in 2009. Measurements of the physical properties (gas reduction factor, transmission of electrons, ...) are ongoing.

Most of the remaining T$_2$ that passes the DPS2-F is trapped in the cryogenic pumping section (CPS) by a layer of Argon frost frozen on the liquid helium cooled beam tube, which forms a highly efficient and large-area cold surface. The accumulation of tritium on the Argon frost increases the potential of tritium migration processes. Therefore a refreshment cycle of the Argon frost is foreseen every 2 months. In order to test the concept of cryo-trapping of T$_2$ on Argon frost a dedicated test experiment was performed. The results showed that a reduction factor of 10$^7$ by the CPS can be achieved \cite{kaz08}. The CPS is presently manufactured and delivery is expected in 2012.

\subsection{Spectrometer section}

The pre-spectrometer as well as the main spectrometer are operated as electrostatic retarding high pass filters. In the final setup the pre-spectrometer will be used as a pre-filter operated on a potential 0~$<$~$U_0$~$<$~E$_0$. A non-zero retarding potential at the pre-spectrometer reduces the flux of $\beta$-electrons into the main-spectrometer and thus reduces the background originating from ionization of residual gas molecules in the volume of the main spectrometer due to $\beta$-electrons. Prior to the installation of the pre-spectrometer into the KATRIN beam line, it was operated in a stand alone test setup.

The purpose of the 24~m long main spectrometer (diameter 10~m) is the high precision energy analysis of the $\beta$-decay electrons. It features an energy resolution of $\Delta E$~=~0.93~eV at 18.6~keV. In order to achieve the desired background rate of $<$~10~mHz, a double layer inner electrode system made of thin wires which is currently being mounted with sub-millimeter precision, is required. The wire layers are put on a more negative potential (100/200~V) with respect to the tank voltage in order to electrostatically shield secondary electrons produced in the vessel wall by cosmic particles. The absolute voltage of $-18.6$~kV needs to be stable on the 1~ppm level and is monitored with a high precision voltage divider \cite{thu09} and an independent monitoring and calibration beam line. The main spectrometer is equipped with an air coil system and a magnetic field sensor system for compensating the Earth's magnetic field and for fine-tuning of the magnetic field close to the analyzing plane.

The installation of the inner electrode system into the main spectrometer is almost finished. The commissioning and electromagnetic test measurements of the main spectrometer are scheduled to start in 2012.

\subsection{Detector system}

The detector system uses two super conducting magnets for guiding the transmitted $\beta$-electrons. The magnet closer to the main spectrometer (pinch magnet) creates a pinch magnetic field of 6~T - which is the largest magnetic field along the KATRIN beam line - and thus defines $\Theta_{max}$ together with $B_S$ (see above). $\beta$-electrons that are able to overcome the potential barriers of the spectrometers are detected in a monolithic 148 pixel silicon PIN\footnote{\textbf{P}ositive \textbf{I}ntrinsic \textbf{N}egative} diode array 90~mm in diameter. The segmentation of the detector is needed to reduce a broadening of the energy resolution caused by inhomogeneities of magnetic field and electric retarding potential in the analyzing plane of the main spectrometer. Because the actual energy analysis of the $\beta$-electrons is done at the main spectrometer, a moderate detector energy resolution of 600~eV FWHM\footnote{\textbf{F}ull \textbf{W}idth at \textbf{H}alf \textbf{M}aximum} is sufficient for the KATRIN experiment. An extensive quality assurance programme with careful selection of materials, shielding and an active veto are used to keep the intrinsic detector background below 1~mHz \cite{leb10}. In addition a positive post acceleration voltage of up to 30~kV can be applied in order to move the signal peak to an energy region of low intrinsic detector background.

After successfully finishing the first commissioning phase of the detector system at the University of Washington in Seattle, the detector system was shipped to KIT. At the moment the detector system is integrated into the KATRIN beam line.

\section{Spectrometer background processes}

\begin{figure*}[hb]
	\centering
	\includegraphics[width=135mm]{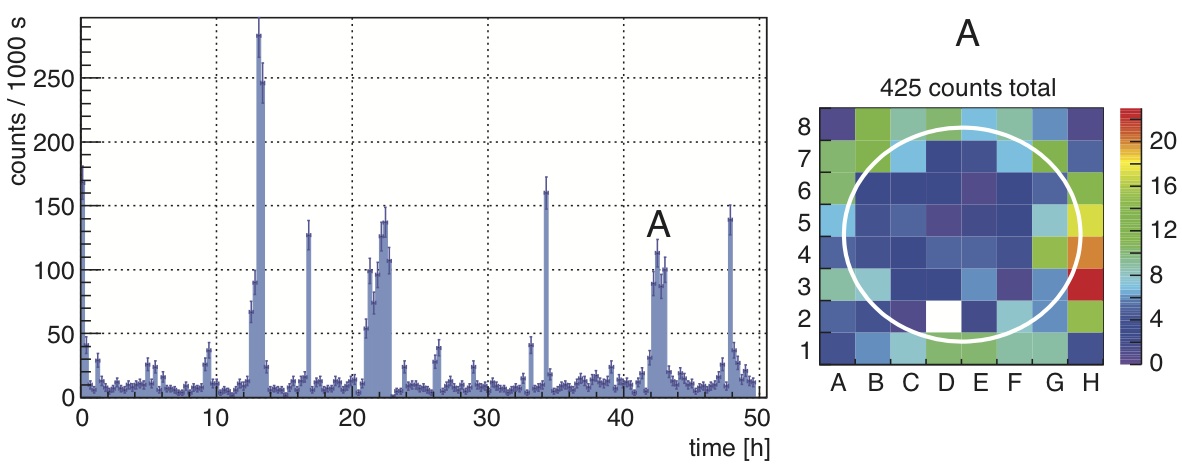}
	\caption{Rate over time plot for a pre-spectrometer background measurement. Each histogram bin has a width of 1000~s. The pixel distribution of an interval of elevated rate (A) is shown on the right side.} \label{fig:radon1}
\end{figure*}

Dedicated background measurements at the pre-spectrometer test setup showed that the decay of radon in the volume of the spectrometer is a major background source \cite{fra10, fra11}: Radon (Rn) atoms, which emanate from materials inside the vacuum region of the KATRIN spectrometers, are able to penetrate deep into the magnetic flux tube so that the final $\alpha$-decay of Rn contributes to the background. Of particular importance are electrons emitted in processes accompanying the Rn $\alpha$-decay such as shake off, internal conversion of excited levels in the Rn daughter atoms, and Auger electrons. While low-energy electrons directly contribute to the background in the signal region, high-energy electrons can be stored magnetically inside the volume of the spectrometer. Depending on their initial energy, they are able to create thousands of secondary electrons via subsequent ionization processes of residual gas molecules thus creating a time dependent background rate at the detector (see figure \ref{fig:radon1}). Due to the magnetron drift caused by the inhomogeneous magnetic field inside the spectrometer, the trapped primary electron moves on a circle therefore creating a ring-like structure of secondary electrons on the detector (see right side of figure \ref{fig:radon1}). For the pre-spectrometer test setup an average Rn induced background rate of 27~$\pm$~6 mHz was determined. The emanation of $^{219}$Rn from the getter material was determined to be 7.5~$\pm$~1.8 mBq thus being responsible for a large fraction (19~$\pm$~4 mHz) of the average background rate. Based on the results of the pre-spectrometer, the background rate from $^{219}$Rn - emanating from 3~km of getter strips - at the main spectrometer is estimated to $\approx$~300~mHz. This is a factor of 30 above the tolerable background and therefore different techniques to reduce the Rn induced background were tested at the pre-spectrometer test setup. A liquid nitrogen (LN2) cooled baffle between getter pump and spectrometer volume could prevent all\footnote{within the uncertainties of the measurement} Rn emanating from the getter material from entering the spectrometer. Another technique that was tested is the removal of stored primary electrons from the Rn decay by applying an electron cyclotron resonance (ECR) pulse with a duration on the order of 100~ms to a spectrometer electrode. The ECR pulse increases the kinetic energy of the stored electron until its cyclotron radius is larger than the spectrometer radius and thus the electron hits the spectrometer wall. Measurements at the pre-spectrometer showed that in principal this technique is working and a background suppression of up to a factor five could be achieved. Since the LN2 cooled baffle showed good results at the pre-spectrometer, also the main spectrometer will be equipped with LN2 cooled baffles in order to achieve the background goal of $<$~10~mHz.

\section{Summary and outlook}

The KATRIN experiment, which is presently being set up at Karlsruhe Institute of Technology (Germany), aims to measure the neutrino mass with a sensitivity of 200~meV/c$^2$ by investigating the kinematics of tritium $\beta$-decay. In order to achieve this sensitivity, the background rate at the detector should not exceed a value of 10~mHz. The installation of the main spectrometer electrode system is almost finished. The detector system - which is the major US contribution to the KATRIN experiment - has been shipped to KIT and is presently being installed. Measurements at the pre-spectrometer test setup showed that Rn decays in the spectrometer volume are a major background source. Different techniques to mitigate Rn induced background were successfully tested at the pre-spectrometer test setup and will be applied to the main spectrometer.

The commissioning and first test measurements at the main spectrometer will start in 2012. Afterwards the complete KATRIN beam line will be commissioned and the neutrino mass measurements will start.


\bigskip 

\end{document}